\documentclass[twocolumn,english,prl]{revtex4}

\usepackage{amssymb}  
\usepackage{graphicx,subfigure}
\usepackage{exscale}
\usepackage{textcomp}
\usepackage{enumerate}
\usepackage{amsmath}
\usepackage{amssymb}
\usepackage{orcidlink}
\usepackage{units}
\usepackage{slashed}
\usepackage{bm}   

\usepackage{xcolor}

\usepackage[normalem]{ulem}  

\newcommand\e{\epsilon}

\newcommand{\ga}{\gamma}
\newcommand{\Ga}{\Gamma}
\newcommand{\de}{\delta}

\newcommand{\ep}{\varepsilon}

\newcommand{\la}{\lambda}

\newcommand{\si}{\sigma}

\renewcommand{\th}{\theta}   

\newcommand{\non}{\nonumber}

\newcommand{\beq}{\begin{equation}}
\newcommand{\eeq}{\end{equation}}
\newcommand{\bal}{\begin{align}}
\newcommand{\eal}{\end{align}}

\newcommand{\ba}[1]{\begin{array}{#1}}
\newcommand{\ea}{\end{array}}

\newcommand{\eqn}[1]{(\ref{#1})}
\newcommand{\sliver}{\kern 0.07em} 

\newcommand{\MeV}{\text{MeV}}

\newcommand{\nsat}{\ensuremath{n_0}}
\newcommand{\ndU}{\ensuremath{n_\text{dUrca}}}
\newcommand{\nue}{{\nu_e}}
\newcommand{\nuebar}{{\bar\nu_e}}

\newcommand{\tn}{\text{n}}  
\newcommand{\tp}{\text{p}}  
\newcommand{\vk}{\mathbf{k}}

\begin{document}

\title{Beyond modified Urca: the nucleon width approximation for flavor-changing processes in dense matter}

\author{Mark G.~Alford\,\orcidlink{0000-0001-9675-7005}}
\email{alford@wustl.edu}
\affiliation{Physics Department, Washington University in Saint Louis, 63130 Saint Louis, MO, USA}

\author{Alexander Haber\,\orcidlink{0000-0002-5511-9565}}
\email{ahaber@physics.wustl.edu}
\affiliation{Physics Department, Washington University in Saint Louis, 63130 Saint Louis, MO, USA}

\author{Ziyuan Zhang\,\orcidlink{0000-0003-4795-0882}}
\email{ziyuan.z@wustl.edu}
\affiliation{Physics Department \& McDonnell Center for the Space Sciences, Washington University in Saint Louis, 63130 Saint Louis, MO, USA}

\date{June 19, 2024}   

\begin{abstract}
Flavor-changing charged current (``Urca'') processes are of central importance in the astrophysics of neutron stars. Standard calculations approximate the Urca rate as the sum of two contributions, direct Urca and modified Urca. Attempts to make modified Urca calculations more accurate have been impeded by an unphysical divergence at the direct Urca threshold density.
In this paper we describe a systematically improvable approach where, in the simplest approximation, instead of modified Urca we include an imaginary part of the nucleon mass (nucleon width). 
The total Urca rate is then obtained via a straightforward generalization of the direct Urca calculation, yielding results that agree with both direct and modified Urca at the densities where those approximations are valid. At low densities, we observe an enhancement of the rate by more than an order of magnitude, with important ramifications for neutron star cooling and other transport properties.

\end{abstract}

\maketitle

\section{Introduction}
\label{sec:intro}
The emission and absorption of neutrinos via Urca (charged-current neutrino-nucleon) processes plays a crucial role in the formation \cite{Janka:2012wk} and thermal evolution \cite{Prakash:1996xs,Page:2005fq} of neutron stars, and in  neutrino transport and proton fraction equilibration in supernovas and neutron star mergers \cite{Alford:2021lpp,Foucart:2022bth, Most:2022yhe, Alford:2023gxq,Espino:2023dei}.

\begin{figure}[t]
\includegraphics[width=\hsize]{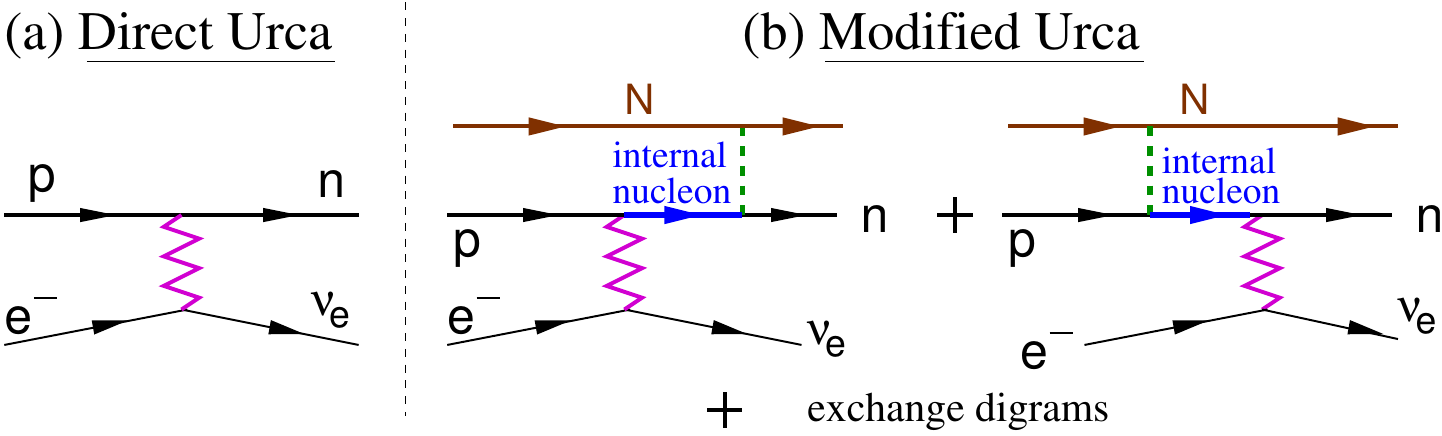}
\caption{Feynman diagrams for direct Urca
and modified Urca contributions to the electron capture  (neutrino creation) rate.
The line labeled N is a spectator nucleon. The green dashed line represents a strong interaction. The modified Urca rate diverges when the internal nucleon (blue) goes on shell.}
\label{fig:dUrca+mUrca}
\end{figure}

At densities and temperatures where neutrinos are trapped and equilibrated the
dominant neutrino creation/absorption mechanism is the direct Urca process  \cite{Lattimer:1991ib}, (Fig.~\ref{fig:dUrca+mUrca}(a)),
\beq
\ba{rl}
\tn &\leftrightarrow \tp\ \ e^-\ \ \nuebar \ , \\
\tn\ \ \nue &\leftrightarrow \tp\ \ e^- \ ,
\ea\label{eq:dUrca-process}
\eeq
However, at lower temperatures where neutrinos are free-streaming,
some nuclear equations of state
have a direct Urca threshold density $\ndU$. At baryon density $n_B$ below $\ndU$ the process \eqn{eq:dUrca-process} is suppressed,
and the standard approach is to add the rate of a separate
 ``modified Urca'' process \cite{Chiu:1964zza}
 \beq
\ba{rl}
\tn\ \ N &\leftrightarrow \tp\ \ N\ \ e^-\ \ \nuebar \ , \\
\tp\ \ N\ \ e^-  &\leftrightarrow \tn\ \ N\ \ \nue  \ ,
\ea\label{eq:mUrca-process}
\eeq
whose Feynman diagrams are shown in Fig.~\ref{fig:dUrca+mUrca}(b).
Here $N$ is a ``spectator'' nucleon, interacting with the participant nucleons via a strong interaction. 
The standard expression for the mUrca rate comes from Friman and Maxwell \cite{Friman:1979ecl}. It has been used widely in the literature \cite{1995A&A...297..717Y, Yakovlev:2000jp}, including calculations of neutron star cooling \cite{Yakovlev:2004iq,Page:2004fy,Ho:2014pta} that are used to constrain the properties of nuclear matter, like the direct Urca threshold \cite{Beloin:2018fyp,Thapa:2022zkr}, or the nuclear superfluid gap \cite{ Beloin:2016zop}.
To obtain an  analytic expression
Friman and Maxwell made many simplifying assumptions such as the Fermi surface approximation (assuming all participating particles are on their Fermi surfaces), neglecting the neutrino momentum, and approximating the propagator of
the internal nucleon (blue line in Fig.~\ref{fig:dUrca+mUrca}(b)) as $1/E_e$.

It would be desirable to have an improvable scheme for calculating Urca processes which would allow us to go beyond some of these approximations. This is needed, for example, to compute flavor relaxation rates in neutron star mergers where temperatures are comparable to the proton Fermi energy so the Fermi surface approximation is not valid, or to compute the absorption mean free path of neutrinos with non-negligible momentum
in matter below the direct Urca threshold density, or to generalize the rate calculation to scenarios such as a high magnetic field.

Improvements on Friman and Maxwell's
calculation  have typically focused on
a better treatment of the
strong interaction with the spectator nucleon (e.g.~\cite{Yakovlev:2000jp,Khodaie:2017fog}; for a review see Ref.~\cite{Schmitt:2017efp}).
However, using a more
accurate representation of the
internal nucleon  propagator leads to
an unphysical divergence in the mUrca rate \cite{Shternin:2018dcn}  as the density approaches $\ndU$ from below.
Currently, the most complete calculation of mUrca processes is that of Suleiman et.\,al.\,\cite{Suleiman:2023bdf} who numerically evaluated the full 10-dimensional phase space integral for the neutrino opacity, but had to introduce a phenomenological infrared cutoff in the charged current correlator \cite{Roberts:2012um,Pascal:2022qeg} to control this divergence.

In this paper, we propose a systematically improvable and practical alternative to the standard approach of calculating dUrca and mUrca as separate rates. This is the {\em nucleon width approximation} (NWA), in which the nucleon masses are given a (density and temperature dependent) imaginary part.
We will focus on nucleonic matter that is degenerate and homogeneous.
We will neglect muons for simplicity, but they can be included in this formalism.  We use natural units where $\hbar=c=k_B=1$.

\section{Urca rates and the charged current correlator}
\label{sec:Urca-cc}

\begin{figure}[t]
\includegraphics[width=0.45\hsize]{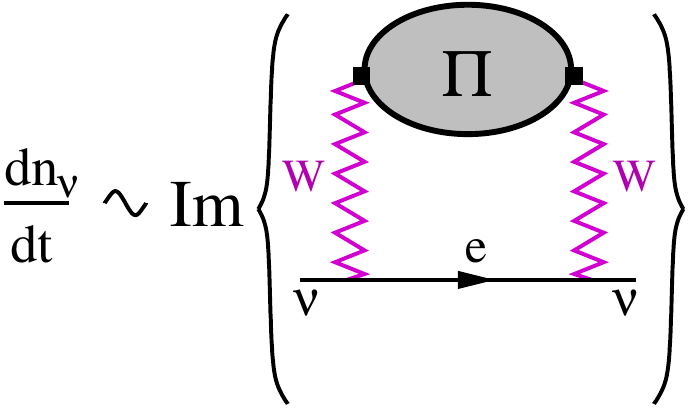}\hspace{0.09\hsize}\includegraphics[width=0.45\hsize]{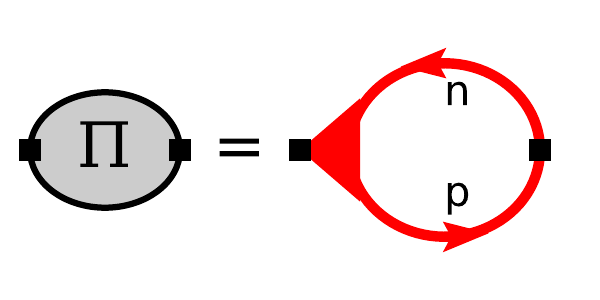}
\caption{Left: Feynman diagram for  imaginary part of neutrino self-energy; the filled ellipse is the hadronic contribution to the in-medium charged current correlator $\Pi$.
Right: Skeleton expansion for $\Pi$ in terms of full  vertex (red triangle) and full nucleon propagators (red lines).}
\label{fig:Urca-unified}
\end{figure}

\begin{figure}
\includegraphics[width=\hsize]{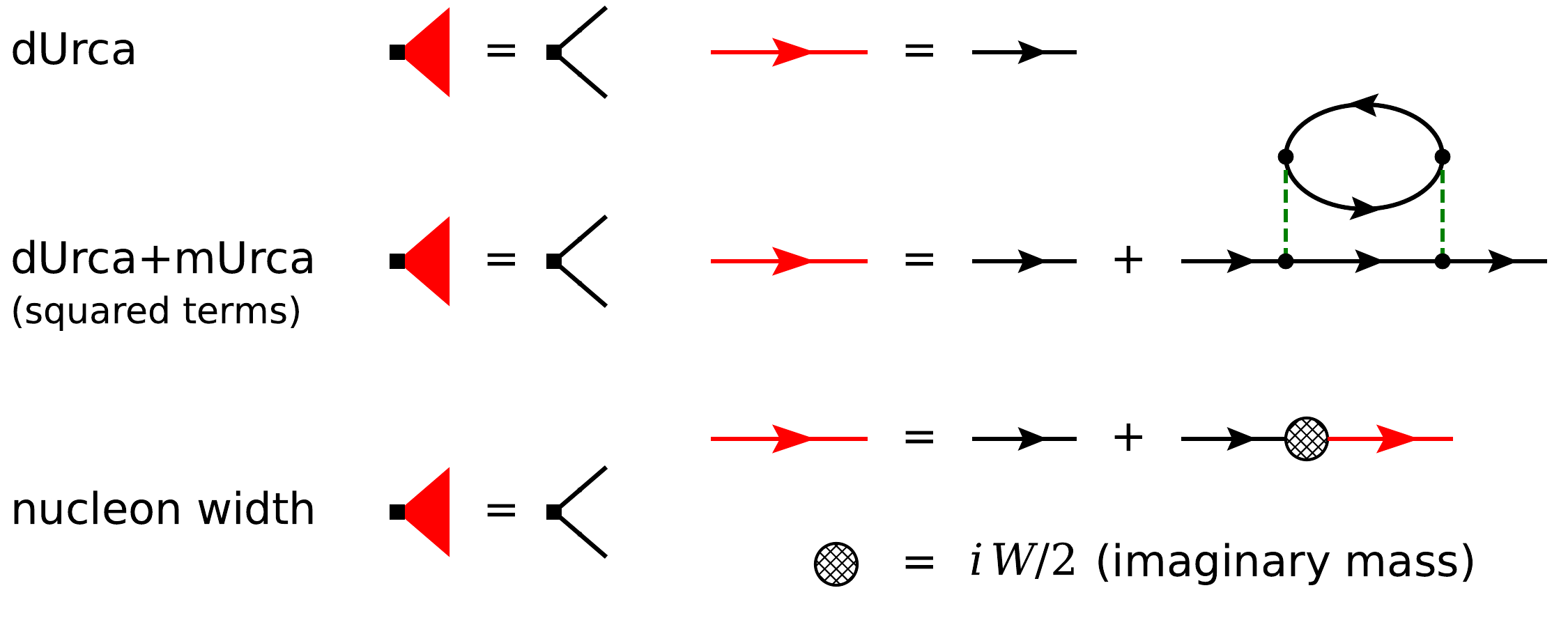}
\caption{
Approximations used in evaluating the in-medium hadronic charged current correlator (Fig.~\ref{fig:Urca-unified}). Dashed lines represent strong interactions. See text for details.
}
\label{fig:Urca-approximations}
\end{figure}

It has been known for some time that the Urca rate 
can be formulated in terms of the imaginary part of the neutrino self-energy \cite{Burrows:1998ek,Sedrakian:2000kc,Lykasov:2008yz,Roberts:2012um,Roberts:2016mwj,Pascal:2022qeg}. This is illustrated in Fig.~\ref{fig:Urca-unified}; according to the Cutkowsky rules the imaginary part can be obtained by cutting the diagram, putting the cut lines on shell, and integrating over their momenta.  The hadronic charged current correlator (W-boson self-energy) $\Pi$ plays a central role: as shown in Fig.~\ref{fig:Urca-unified} it can be written in a skeleton expansion with a full charged current vertex $V^\text{full}$ (red triangle) and one full neutron propagator $G^\text{full}_\tn$ and one full proton propagator $G^\text{full}_\tp$ (thick red lines),
\beq
\Pi_{\la\si}(q) = \int \dfrac{d^4 k}{(2\pi)^4}
\text{Tr}\Bigl[
V^\text{full}_\la G^\text{full}_\tp(k{-}q) V^\text{full}_\si G^\text{full}_\tn(k) \Bigr] \ .
\label{eq:Pi-skeleton}
\eeq
At nonzero temperature the integral over $k_0$ becomes a sum over Matsubara frequencies.
In this framework we can understand the standard approaches to calculating Urca rates as different approximations to these components of the skeleton expansion (Fig.~\ref{fig:Urca-approximations}).\\[1ex]
(1) \underline{Direct Urca} (Fig.~\ref{fig:Urca-approximations}, first row) corresponds to 
evaluating \eqn{eq:Pi-skeleton} in the
approximation where the vertex is derived from the bare charged current, 
\beq
V^W_\mu = \dfrac{g_w \cos\th_C}{2\sqrt{2}} \ga_\mu (1 -g_A\ga_5) \ ,
\label{eq:W-vertex}
\eeq
(which can be generalized to include additional terms such as weak magnetism \cite{Vogel:1983hi,Horowitz:2003yx})
and $G^\text{full}$ is replaced by the mean-field  nucleon propagator, with isospin index $a$
specifying neutron or proton,
\beq 
G_a^\text{mf}(k) = \dfrac{1}{(k_0-U_a) \ga^0 + k_i\ga^i + M^*_a }\ .
\label{eq:prop-undressed} 
\eeq
At finite density one includes a chemical potential in the $\ga^0$ term and an appropriate $i\ep$ prescription.
The mean-field propagators include single-particle in-medium corrections via an effective mass $M^*_a$ and energy shift $U_a$. 
\\[1ex]
(2) \underline{Direct + modified Urca} (Fig.~\ref{fig:Urca-approximations}, second row) is currently the standard approach. It is an approximation using the bare charged current vertex and adding a second term to the nucleon propagator where a model of the strong interaction is used to dress it with a single particle-hole companion (summed over $n$-$\bar n$ and $p$-$\bar p$). By substituting this in to the skeleton expansion of the hadronic charged current correlator $\Pi$ (Fig.~\ref{fig:Urca-unified}) one can see that this is equivalent to summing over the squares of the amplitudes shown in  Fig.~\ref{fig:dUrca+mUrca}(b). 
The interference terms between these diagrams are not included (in Fig.~\ref{fig:Urca-unified} they would correspond to vertex corrections in $\Pi$) but these interference terms are already known to be a small correction \cite{Shternin:2018dcn}. 

As noted above, the mUrca contribution has an unphysical divergence when the nucleon propagator between the charged current vertex and the strong interaction vertex goes on shell.\\[1ex]
(3) \underline{Nucleon Width Approximation (NWA)}. 
A more consistent approach is the nucleon width approximation (Fig.~\ref{fig:Urca-approximations}, bottom row) in which we evaluate \eqn{eq:Pi-skeleton} using the bare vertex \eqn{eq:W-vertex} and
a dressed nucleon propagator that includes an imaginary contribution to the mass.
The full
nucleon propagator can be expressed in terms of the self-energy (hatched circle) via a Schwinger-Dyson equation (Fig.~\ref{fig:Urca-approximations}, third row). 
The self-energy is determined by a model of the strong interaction. In general it would contain a sum of different
Dirac matrix structures \cite{Dieperink:1990kw}, each being a function of the
energy and momentum flowing through it. In the nucleon width approximation we keep
only the imaginary Lorentz scalar component $iW/2$, with no energy or
momentum dependence. The NWA nucleon propagator then has the same form as the mean-field one, with an imaginary part $i W_a/2$ added to the mass
\beq
G_a^\text{NWA}(k) = \dfrac{1}{(k_0-U_a) \ga^0 + k_i\ga^i + M^*_a + iW_a/2} \ .
\label{eq:prop-dressed}
\eeq


It should be noted that this approximation, where there is a nucleon width but no vertex correction, is not appropriate for neutral current processes such as elastic scattering of neutrinos.  The
vector component of the neutral current is the exactly conserved baryon current $J^B_\la$ so the vector-vector component of the neutral current correlator, $\Pi_{BB}^{\la\si}(q)$,  obeys a Ward identity $q_\la\Pi_{BB}^{\la\si}=0$, but giving the nucleon a width without introducing compensating corrections to the vertex will violate this condition (see Ref.~\cite{Peskin:1995ev}, Ch.~7.5 and 21.3).
In contrast, the vector component of the charged current is the non-conserved isospin current \cite{Leinson:2001ei}. Thus the Ward identity for the vector-vector part of the charged current correlator is already violated in vacuum by $M_\tn-M_\tp \sim 1\,\MeV$,  and in 
beta-equilibrated nuclear matter by $M^*_\tn-M^*_\tp$ or $U_\tn-U_\tp$ which can be as large as tens of MeV \cite{Roberts:2012um,Horowitz:2012us}.
The nucleon widths that we will use are of order $T^2/(5\,\MeV)$
and so for $T\lesssim 10\,\MeV$ are no larger than the intrinsic violation of isospin symmetry.



In the nucleon width approximation the total Urca rate $\Ga^\text{NWA}$ takes a similar form to the dUrca approximation, except that in the charged current correlator  we use
nucleon propagators with widths, $G^\text{full}_a\to G^\text{NWA}_a$ (Eq.~\eqn{eq:prop-dressed}). 

The propagator for a fermion with nonzero width
can be written as a mass-spectral decomposition \cite{Kuksa:2015} in terms of propagators with zero width, so in our context 
\beq
G^\text{NWA}_a(k,M^*_a,W_a) = \int_{-\infty}^\infty \!dm\,
G^\text{mf}_a(k,m) \, R_a(m) \ ,
\label{eq:prop-with-width}
\eeq
where the mass-spectral function takes the Breit-Wigner form
\beq
R_a(m) \equiv \dfrac{1}{\pi}\dfrac{W_a/2}{(m-M^*_a)^2 + W_a^2/4} \ .
\label{eq:breit-wigner-R}
\eeq
In the limit where the width $W_a\to 0$, $R_a(m)\to\de(m-M^*_a)$, and 
$G^\text{NWA}_a \to G^\text{mf}_a$.

Substituting \eqn{eq:prop-with-width} into \eqn{eq:Pi-skeleton} we find 
\begin{align}
\Pi^\text{NWA}_{\la\si}(q)
= \int_{-\infty}^\infty & \!\!dm_\tn dm_\tp 
\,\Pi^\text{mf}_{\la\si}(q,m_\tn,m_\tp) 
R_\tn(m_\tn) R_\tp(m_\tp)
\ .  \label{eq:Pi-nw}
\end{align}

Since the Urca rate is just an integral over $q$ of the correlator $\Pi(q)$ multiplied by a function of $q$ that comes from the leptonic part
of the neutrino self-energy diagram (Fig.~\ref{fig:Urca-unified}), and the dUrca rate is obtained by using the mean-field propagators in the correlator,
this leads to a particularly simple form for Urca rates 
in the nucleon-width approximation: one just ``smears'' the dUrca rate over a range of nucleon masses
\beq
\Gamma^\text{NWA}
= \int_{-\infty}^\infty \!\!dm_\tn dm_\tp 
\Gamma^\text{dUrca}(m_\tn,m_\tp)
\,R_\tn(m_\tn) R_\tp(m_\tp)
\ .  \label{eq:Urca-rate-nw}
\eeq


This expression is applicable to any Urca process, i.e. any process that is obtained by cutting the neutrino self-energy diagram (Fig.~\ref{fig:Urca-unified}).
Once the width has been obtained from a model of the strong interaction, or by a phenomenological fit,
the total Urca rate can be straightforwardly calculated from the dUrca rate for general nucleon masses.

\begin{figure}
\includegraphics[width=0.95\hsize]{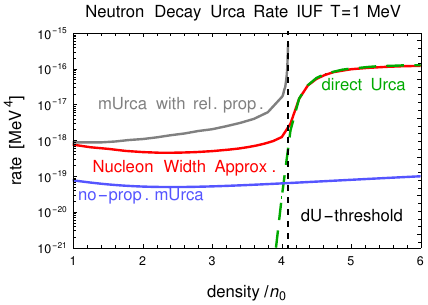}
\caption{Neutron decay rate as a function of density for IUF matter at $T=1\,\MeV$, comparing NWA with standard approximations. See text for details.
}
\label{fig:Urca_rates}
\end{figure}

\begin{figure}
  \centering
  \includegraphics[width=0.95\hsize]{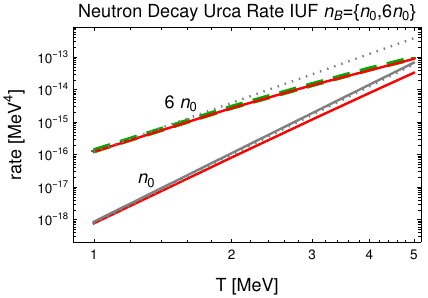}
  \caption{
  Neutron decay rate as a function of temperature for IUF matter at densities $n_B=\nsat \equiv 0.16\,\text{fm}^{-3}$ (well below the dUrca threshold)
   and $n_B=6\, \nsat$ (above threshold), showing that NWA gives the expected temperature dependence; see text for details. 
}
\label{fig:Tdep}
\end{figure}

\begin{figure}
\includegraphics[width=0.85\hsize]{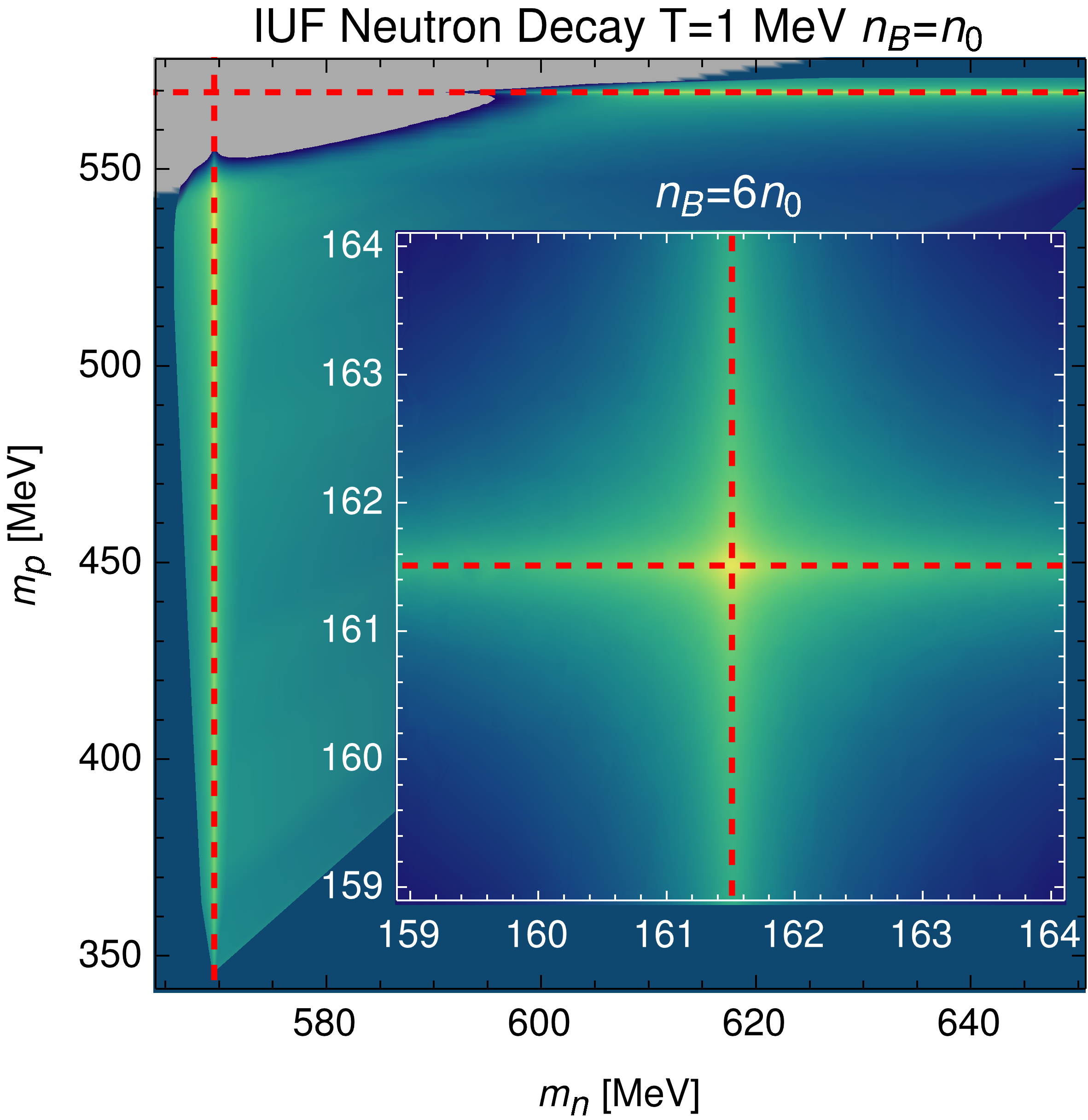}
\caption{The NWA Urca rate integrand \eqn{eq:Urca-rate-nw} for the IUF EoS. Brighter (yellow) color indicates a larger integrand. Main plot is for a density below the dUrca threshold, insert is above the threshold. Dashed red lines show the effective masses $M^*_a$. In the outer blue-colored area we do not calculate the integrand because its contribution to the integrand is quadratically suppressed by the Breit-Wigner functions. In the gray-shaded area, the integrand is exponentially suppressed due to kinematic constraints.
}\label{fig:heat_plot}
\end{figure}

\section{Results}
\label{sec:results}

In Fig.~\ref{fig:Urca_rates} we show the density dependence of the neutron decay rate in matter described by the IUF equation of state \cite{Fattoyev:2010mx} at $T=1\,\MeV$ in cold chemical equilibrium $\mu_n=\mu_p+\mu_e$ \cite{Alford:2018lhf, Alford:2021ogv, Alford:2023gxq}. The NWA calculation (red line) uses nucleon widths $W_a=T^2/(5\,\MeV)$ obtained from a Brueckner theory calculation for pure neutron matter  using the Paris NN potential (Ref.~\cite{Sedrakian:2000kc}, Eq.~(69)), which found only a weak density dependence of the width, justifying the assumptions for NWA. One could explore other models of the strong interaction, and use their estimates for the widths, or alternatively set the width phenomenologically by matching $\Ga^\text{NWA}$ to an mUrca-with-propagator calculation at a density well below the dUrca threshold.  

In  Fig.~\ref{fig:Urca_rates}
the dUrca rate is calculated by evaluating the full phase space integral and using
the relativistic matrix element as in Ref.~\cite{Alford:2023gxq}.
We see that the NWA result agrees with dUrca above the dUrca threshold. 
Far below the threshold, it is fairly close to 
an improved mUrca calculation (gray line), where, as in Ref.~\cite{Shternin:2018dcn},  the internal
nucleon propagator is included. This improved mUrca calculation
uses the Fermi surface approximation and models the strong interaction via one-pion exchange as in Ref.~\cite{Yakovlev:2000jp}, but with relativistic kinematics and propagators for the nucleons.
As expected,
the improved mUrca rate diverges at the dUrca threshold, while NWA smoothly matches to dUrca.

In Fig.~\ref{fig:Tdep} we show the temperature dependence of the neutron decay rate for IUF matter at two densities.
Firstly $n_B=\nsat$ (saturation density), far below the dUrca threshold. Secondly $n_B=6\nsat$, above the dUrca threshold. The NWA calculation uses the same width estimates as in Fig.~\ref{fig:Urca_rates}. We see that above threshold NWA (solid red line) agrees very well with the dUrca calculation (dashed green line). 

Far below threshold, NWA  (solid red line) agrees well with the improved mUrca calculation (solid grey line, for details see description of Fig.~\ref{fig:Urca_rates}).
In both cases, the dotted lines show the predicted power-law behavior ($T^5$ for dUrca, $T^7$ for mUrca) at low temperatures where the Fermi surface approximation can be used to simplify the integrals \cite{Friman:1979ecl,Yakovlev:2000jp}.  We see that NWA captures the expected temperature dependence above and below the threshold.

From Figs.~\ref{fig:Urca_rates} and \ref{fig:Tdep} we see that NWA predicts that the rate at low densities is enhanced by at least an order of magnitude compared to the widely used no-propagator mUrca calculation (blue line in Fig.~\ref{fig:Urca_rates}). Such an enhancement was also found in the improved mUurca calculation of Ref.~\cite{Shternin:2018dcn}, but there it was accompanied by a divergence at the dUrca threshold. In NWA we see the enhancement in a well-behaved calculation that can be applied aross a wide range of densities and temperatures. This may have important ramifications for
any scenario where charged current neutrino interactions play an important role, such as neutron star cooling, transport in neutron stars and neutron star mergers \cite{Yakovlev:2000jp, Gavassino:2020kwo, Radice:2021jtw, Foucart:2022bth, Gavassino:2023xkt} and  for the thermal states of compact stars in
low-mass X-ray binaries \cite{Yakovlev:2004iq,Fortin:2017rxq,Shternin:2018dcn}.

To understand how NWA
implements the physics that the modified Urca process attempts to capture, we show in Fig.~\ref{fig:heat_plot} the integrand from
\eqn{eq:Urca-rate-nw} plotted in the $(m_n,m_p)$ plane.
At densities below the direct Urca threshold (Fig.~\ref{fig:heat_plot}, main plot) the direct Urca rate $\Gamma^\text{dUrca}(M^*_\tn,M^*_\tp)$ is exponentially suppressed in the $T\to 0$ limit because $(M^*_\tn,M^*_\tp)$ (intersection of dashed red lines) lies outside the kinematically allowed region. Specifically,
the proton and electron Fermi momenta are too small to produce a neutron on its Fermi surface, violating the dUrca criterion
$k_{F\tp} + k_{Fe} \geq k_{F\tn}$.
But in the NWA integral \eqn{eq:Urca-rate-nw} there are contributions from lower $m_p$ (or higher $m_n$) (bright regions)  where, since the chemical potential is held constant in the spectral mass integral, the proton Fermi momentum is larger, or neutron Fermi momentum is smaller, and hence obeys the dUrca criterion. 
These contributions are moderately suppressed because they lie
in the tails of the Breit-Wigner distribution \eqn{eq:breit-wigner-R},
yielding the slower (than dUrca) rate that is seen in the
improved mUrca calculation.

At a density above the direct Urca threshold (inset in Fig.~\ref{fig:heat_plot}) the in-medium masses $(M^*_n,M^*_p)$ enter the kinematically allowed region.
The Breit-Wigner function is then effectively just a slightly smeared delta function, and the mass integral yields almost the same result as setting $m_a=M_a^*$ in the rate. Thus instead of the unphysical divergence seen in mUrca calculations we obtain a smooth crossover to the standard dUrca rate.

\section{Conclusions}
\label{sec:conclusions}

We have argued that the nucleon width approximation (NWA) is a convenient method for
calculating Urca rates that, unlike the dUrca+mUrca approximation,
represents the first step in a systematically improvable scheme. The rate calculated using NWA, with widths taken from a Brueckner theory calculation,  shows the expected behavior above and below the dUrca threshold and varies smoothly across it. 
The rate below threshold is enhanced by an order of magnitude compared to standard mUrca calculations.
 
NWA can be applied in any
context where dUrca rates can be calculated, opening the door to a consistent study of total Urca rates at finite temperature, which has not been possible up to this point, and calculations of total Urca rates for matter with non-equilibrium neutrino distributions, strong magnetic fields, or other scenarios outside the scope of standard mUrca calculations like decays in some models of dark matter \cite{Fornal:2018eol,Husain:2022bxl,Shirke:2023ktu,Shirke:2024ymc}, weak decays in hyperonic matter \cite{Haensel:2001em,vanDalen:2003uy,Ofengeim:2019fjy, Alford:2020pld, Xu:2021jns}, or weak processes in quark matter \cite{ Jaikumar:2005hy,Berdermann:2016mwt,Hernandez:2024rxi}. 

For our calculations we evaluated the dUrca rate by performing the full 4D numerical phase space integral with the vacuum matrix element \cite{Alford:2023gxq} so the NWA rate is a 6D numerical integral which is tractable because the integrand is well behaved. At low temperatures ($T\lesssim 1\,\MeV$) one could alternatively use an analytic approximation for the dUrca rate
based on the Fermi surface approximation with either a constant \cite{Yakovlev:2000jp} or an angle-averaged matrix element. These approaches agree with the full integral to within a factor of about $2-7$ at low $T$ depending on what approximation is used for the matrix element. Then the NWA rate becomes a 2D numerical integral of a well-behaved peaked integrand as shown in Fig.~\ref{fig:heat_plot}.

As well as investigating applications of NWA  (to neutron star cooling for example), 
a natural next step would be to pursue the systematic improvement scheme outlined in this letter: (1) explore other estimates of the nucleon width, e.g. from chiral effective theory \cite{Vidana:2022ket}; (2) include other contributions (with different Dirac index structures) to the nucleon self-energy; (3) allow for momentum and energy dependence of the self-energy; (4) include vertex corrections, for example an RPA resummation along the lines of Ref.~\cite{Shin:2023sei}.

\section{Acknowledgements}
We thank S. Reddy, G. Baym, A. Sedrakian, and S. Harris for helpful discussions.
MGA thanks the Yukawa Institute for Theoretical Physics at Kyoto University and RIKEN iTHEMS for the workshop (YITP-T-23-05) on “Condensed Matter Physics of QCD 2024” which provided useful discussions for this work.
MGA and AH are partly supported by the U.S. Department of Energy, Office of Science, Office of Nuclear Physics, under Award No.~\#DE-FG02-05ER41375. ZZ is supported in part by the National Science Foundation (NSF) within the framework of the MUSES collaboration, under grant number OAC-2103680.

\bibliography{medium_urca}

\begin{thebibliography}{57}
\expandafter\ifx\csname natexlab\endcsname\relax\def\natexlab#1{#1}\fi
\expandafter\ifx\csname bibnamefont\endcsname\relax
  \def\bibnamefont#1{#1}\fi
\expandafter\ifx\csname bibfnamefont\endcsname\relax
  \def\bibfnamefont#1{#1}\fi
\expandafter\ifx\csname citenamefont\endcsname\relax
  \def\citenamefont#1{#1}\fi
\expandafter\ifx\csname url\endcsname\relax
  \def\url#1{\texttt{#1}}\fi
\expandafter\ifx\csname urlprefix\endcsname\relax\def\urlprefix{URL }\fi
\providecommand{\bibinfo}[2]{#2}
\providecommand{\eprint}[2][]{\url{#2}}

\bibitem[{\citenamefont{Janka}(2012)}]{Janka:2012wk}
\bibinfo{author}{\bibfnamefont{H.-T.} \bibnamefont{Janka}},
  \bibinfo{journal}{Ann. Rev. Nucl. Part. Sci.} \textbf{\bibinfo{volume}{62}},
  \bibinfo{pages}{407} (\bibinfo{year}{2012}), \eprint{1206.2503}.

\bibitem[{\citenamefont{Prakash et~al.}(1997)\citenamefont{Prakash, Bombaci,
  Prakash, Ellis, Lattimer, and Knorren}}]{Prakash:1996xs}
\bibinfo{author}{\bibfnamefont{M.}~\bibnamefont{Prakash}},
  \bibinfo{author}{\bibfnamefont{I.}~\bibnamefont{Bombaci}},
  \bibinfo{author}{\bibfnamefont{M.}~\bibnamefont{Prakash}},
  \bibinfo{author}{\bibfnamefont{P.~J.} \bibnamefont{Ellis}},
  \bibinfo{author}{\bibfnamefont{J.~M.} \bibnamefont{Lattimer}},
  \bibnamefont{and} \bibinfo{author}{\bibfnamefont{R.}~\bibnamefont{Knorren}},
  \bibinfo{journal}{Phys. Rept.} \textbf{\bibinfo{volume}{280}},
  \bibinfo{pages}{1} (\bibinfo{year}{1997}), \eprint{nucl-th/9603042}.

\bibitem[{\citenamefont{Page et~al.}(2006)\citenamefont{Page, Geppert, and
  Weber}}]{Page:2005fq}
\bibinfo{author}{\bibfnamefont{D.}~\bibnamefont{Page}},
  \bibinfo{author}{\bibfnamefont{U.}~\bibnamefont{Geppert}}, \bibnamefont{and}
  \bibinfo{author}{\bibfnamefont{F.}~\bibnamefont{Weber}},
  \bibinfo{journal}{Nucl. Phys. A} \textbf{\bibinfo{volume}{777}},
  \bibinfo{pages}{497} (\bibinfo{year}{2006}), \eprint{astro-ph/0508056}.

\bibitem[{\citenamefont{Alford et~al.}(2021{\natexlab{a}})\citenamefont{Alford,
  Harutyunyan, and Sedrakian}}]{Alford:2021lpp}
\bibinfo{author}{\bibfnamefont{M.}~\bibnamefont{Alford}},
  \bibinfo{author}{\bibfnamefont{A.}~\bibnamefont{Harutyunyan}},
  \bibnamefont{and}
  \bibinfo{author}{\bibfnamefont{A.}~\bibnamefont{Sedrakian}},
  \bibinfo{journal}{Phys. Rev. D} \textbf{\bibinfo{volume}{104}},
  \bibinfo{pages}{103027} (\bibinfo{year}{2021}{\natexlab{a}}),
  \eprint{2108.07523}.

\bibitem[{\citenamefont{Foucart}(2023)}]{Foucart:2022bth}
\bibinfo{author}{\bibfnamefont{F.}~\bibnamefont{Foucart}},
  \bibinfo{journal}{Liv. Rev. Comput. Astrophys.} \textbf{\bibinfo{volume}{9}},
  \bibinfo{pages}{1} (\bibinfo{year}{2023}), \eprint{2209.02538}.

\bibitem[{\citenamefont{Most et~al.}(2024)\citenamefont{Most, Haber, Harris,
  Zhang, Alford, and Noronha}}]{Most:2022yhe}
\bibinfo{author}{\bibfnamefont{E.~R.} \bibnamefont{Most}},
  \bibinfo{author}{\bibfnamefont{A.}~\bibnamefont{Haber}},
  \bibinfo{author}{\bibfnamefont{S.~P.} \bibnamefont{Harris}},
  \bibinfo{author}{\bibfnamefont{Z.}~\bibnamefont{Zhang}},
  \bibinfo{author}{\bibfnamefont{M.~G.} \bibnamefont{Alford}},
  \bibnamefont{and} \bibinfo{author}{\bibfnamefont{J.}~\bibnamefont{Noronha}},
  \bibinfo{journal}{Astrophys. J. Lett.} \textbf{\bibinfo{volume}{967}},
  \bibinfo{pages}{L14} (\bibinfo{year}{2024}), \eprint{2207.00442}.

\bibitem[{\citenamefont{Alford et~al.}(2024)\citenamefont{Alford, Haber, and
  Zhang}}]{Alford:2023gxq}
\bibinfo{author}{\bibfnamefont{M.~G.} \bibnamefont{Alford}},
  \bibinfo{author}{\bibfnamefont{A.}~\bibnamefont{Haber}}, \bibnamefont{and}
  \bibinfo{author}{\bibfnamefont{Z.}~\bibnamefont{Zhang}},
  \bibinfo{journal}{Phys. Rev. C} \textbf{\bibinfo{volume}{109}},
  \bibinfo{pages}{055803} (\bibinfo{year}{2024}), \eprint{2306.06180}.

\bibitem[{\citenamefont{Espino et~al.}(2024)\citenamefont{Espino, Hammond,
  Radice, Bernuzzi, Gamba, Zappa, Longo~Micchi, and Perego}}]{Espino:2023dei}
\bibinfo{author}{\bibfnamefont{P.~L.} \bibnamefont{Espino}},
  \bibinfo{author}{\bibfnamefont{P.}~\bibnamefont{Hammond}},
  \bibinfo{author}{\bibfnamefont{D.}~\bibnamefont{Radice}},
  \bibinfo{author}{\bibfnamefont{S.}~\bibnamefont{Bernuzzi}},
  \bibinfo{author}{\bibfnamefont{R.}~\bibnamefont{Gamba}},
  \bibinfo{author}{\bibfnamefont{F.}~\bibnamefont{Zappa}},
  \bibinfo{author}{\bibfnamefont{L.~F.} \bibnamefont{Longo~Micchi}},
  \bibnamefont{and} \bibinfo{author}{\bibfnamefont{A.}~\bibnamefont{Perego}},
  \bibinfo{journal}{Phys. Rev. Lett.} \textbf{\bibinfo{volume}{132}},
  \bibinfo{pages}{211001} (\bibinfo{year}{2024}), \eprint{2311.00031}.

\bibitem[{\citenamefont{Lattimer et~al.}(1991)\citenamefont{Lattimer, Prakash,
  Pethick, and Haensel}}]{Lattimer:1991ib}
\bibinfo{author}{\bibfnamefont{J.~M.} \bibnamefont{Lattimer}},
  \bibinfo{author}{\bibfnamefont{M.}~\bibnamefont{Prakash}},
  \bibinfo{author}{\bibfnamefont{C.~J.} \bibnamefont{Pethick}},
  \bibnamefont{and} \bibinfo{author}{\bibfnamefont{P.}~\bibnamefont{Haensel}},
  \bibinfo{journal}{Phys. Rev. Lett.} \textbf{\bibinfo{volume}{66}},
  \bibinfo{pages}{2701} (\bibinfo{year}{1991}).

\bibitem[{\citenamefont{Chiu and Salpeter}(1964)}]{Chiu:1964zza}
\bibinfo{author}{\bibfnamefont{H.-Y.} \bibnamefont{Chiu}} \bibnamefont{and}
  \bibinfo{author}{\bibfnamefont{E.~E.} \bibnamefont{Salpeter}},
  \bibinfo{journal}{Phys. Rev. Lett.} \textbf{\bibinfo{volume}{12}},
  \bibinfo{pages}{413} (\bibinfo{year}{1964}).

\bibitem[{\citenamefont{Friman and Maxwell}(1979)}]{Friman:1979ecl}
\bibinfo{author}{\bibfnamefont{B.~L.} \bibnamefont{Friman}} \bibnamefont{and}
  \bibinfo{author}{\bibfnamefont{O.~V.} \bibnamefont{Maxwell}},
  \bibinfo{journal}{Astrophys. J.} \textbf{\bibinfo{volume}{232}},
  \bibinfo{pages}{541} (\bibinfo{year}{1979}).

\bibitem[{\citenamefont{{Yakovlev} and
  {Levenfish}}(1995)}]{1995A&A...297..717Y}
\bibinfo{author}{\bibfnamefont{D.~G.} \bibnamefont{{Yakovlev}}}
  \bibnamefont{and} \bibinfo{author}{\bibfnamefont{K.~P.}
  \bibnamefont{{Levenfish}}}, \bibinfo{journal}{Astron.~\& Astrophys.}
  \textbf{\bibinfo{volume}{297}}, \bibinfo{pages}{717} (\bibinfo{year}{1995}).

\bibitem[{\citenamefont{Yakovlev et~al.}(2001)\citenamefont{Yakovlev, Kaminker,
  Gnedin, and Haensel}}]{Yakovlev:2000jp}
\bibinfo{author}{\bibfnamefont{D.~G.} \bibnamefont{Yakovlev}},
  \bibinfo{author}{\bibfnamefont{A.~D.} \bibnamefont{Kaminker}},
  \bibinfo{author}{\bibfnamefont{O.~Y.} \bibnamefont{Gnedin}},
  \bibnamefont{and} \bibinfo{author}{\bibfnamefont{P.}~\bibnamefont{Haensel}},
  \bibinfo{journal}{Phys. Rept.} \textbf{\bibinfo{volume}{354}},
  \bibinfo{pages}{1} (\bibinfo{year}{2001}), \eprint{astro-ph/0012122}.

\bibitem[{\citenamefont{Yakovlev and Pethick}(2004)}]{Yakovlev:2004iq}
\bibinfo{author}{\bibfnamefont{D.~G.} \bibnamefont{Yakovlev}} \bibnamefont{and}
  \bibinfo{author}{\bibfnamefont{C.~J.} \bibnamefont{Pethick}},
  \bibinfo{journal}{Ann. Rev. Astron. Astrophys.}
  \textbf{\bibinfo{volume}{42}}, \bibinfo{pages}{169} (\bibinfo{year}{2004}),
  \eprint{astro-ph/0402143}.

\bibitem[{\citenamefont{Page et~al.}(2004)\citenamefont{Page, Lattimer,
  Prakash, and Steiner}}]{Page:2004fy}
\bibinfo{author}{\bibfnamefont{D.}~\bibnamefont{Page}},
  \bibinfo{author}{\bibfnamefont{J.~M.} \bibnamefont{Lattimer}},
  \bibinfo{author}{\bibfnamefont{M.}~\bibnamefont{Prakash}}, \bibnamefont{and}
  \bibinfo{author}{\bibfnamefont{A.~W.} \bibnamefont{Steiner}},
  \bibinfo{journal}{Astrophys. J. Suppl.} \textbf{\bibinfo{volume}{155}},
  \bibinfo{pages}{623} (\bibinfo{year}{2004}), \eprint{astro-ph/0403657}.

\bibitem[{\citenamefont{Ho et~al.}(2015)\citenamefont{Ho, Elshamouty, Heinke,
  and Potekhin}}]{Ho:2014pta}
\bibinfo{author}{\bibfnamefont{W.~C.~G.} \bibnamefont{Ho}},
  \bibinfo{author}{\bibfnamefont{K.~G.} \bibnamefont{Elshamouty}},
  \bibinfo{author}{\bibfnamefont{C.~O.} \bibnamefont{Heinke}},
  \bibnamefont{and} \bibinfo{author}{\bibfnamefont{A.~Y.}
  \bibnamefont{Potekhin}}, \bibinfo{journal}{Phys. Rev. C}
  \textbf{\bibinfo{volume}{91}}, \bibinfo{pages}{015806}
  (\bibinfo{year}{2015}), \eprint{1412.7759}.

\bibitem[{\citenamefont{Beloin et~al.}(2019)\citenamefont{Beloin, Han, Steiner,
  and Odbadrakh}}]{Beloin:2018fyp}
\bibinfo{author}{\bibfnamefont{S.}~\bibnamefont{Beloin}},
  \bibinfo{author}{\bibfnamefont{S.}~\bibnamefont{Han}},
  \bibinfo{author}{\bibfnamefont{A.~W.} \bibnamefont{Steiner}},
  \bibnamefont{and}
  \bibinfo{author}{\bibfnamefont{K.}~\bibnamefont{Odbadrakh}},
  \bibinfo{journal}{Phys. Rev. C} \textbf{\bibinfo{volume}{100}},
  \bibinfo{pages}{055801} (\bibinfo{year}{2019}), \eprint{1812.00494}.

\bibitem[{\citenamefont{Thapa and Sinha}(2022)}]{Thapa:2022zkr}
\bibinfo{author}{\bibfnamefont{V.~B.} \bibnamefont{Thapa}} \bibnamefont{and}
  \bibinfo{author}{\bibfnamefont{M.}~\bibnamefont{Sinha}}
  (\bibinfo{year}{2022}), \eprint{2203.02272}.

\bibitem[{\citenamefont{Beloin et~al.}(2018)\citenamefont{Beloin, Han, Steiner,
  and Page}}]{Beloin:2016zop}
\bibinfo{author}{\bibfnamefont{S.}~\bibnamefont{Beloin}},
  \bibinfo{author}{\bibfnamefont{S.}~\bibnamefont{Han}},
  \bibinfo{author}{\bibfnamefont{A.~W.} \bibnamefont{Steiner}},
  \bibnamefont{and} \bibinfo{author}{\bibfnamefont{D.}~\bibnamefont{Page}},
  \bibinfo{journal}{Phys. Rev. C} \textbf{\bibinfo{volume}{97}},
  \bibinfo{pages}{015804} (\bibinfo{year}{2018}), \eprint{1612.04289}.

\bibitem[{\citenamefont{Khodaie et~al.}(2017)\citenamefont{Khodaie,
  Dehghan~Niri, Moshfegh, and Haensel}}]{Khodaie:2017fog}
\bibinfo{author}{\bibfnamefont{A.}~\bibnamefont{Khodaie}},
  \bibinfo{author}{\bibfnamefont{A.}~\bibnamefont{Dehghan~Niri}},
  \bibinfo{author}{\bibfnamefont{H.~R.} \bibnamefont{Moshfegh}},
  \bibnamefont{and} \bibinfo{author}{\bibfnamefont{P.}~\bibnamefont{Haensel}},
  \bibinfo{journal}{Acta Phys. Polon. B} \textbf{\bibinfo{volume}{48}},
  \bibinfo{pages}{661} (\bibinfo{year}{2017}).

\bibitem[{\citenamefont{Schmitt and Shternin}(2018)}]{Schmitt:2017efp}
\bibinfo{author}{\bibfnamefont{A.}~\bibnamefont{Schmitt}} \bibnamefont{and}
  \bibinfo{author}{\bibfnamefont{P.}~\bibnamefont{Shternin}},
  \bibinfo{journal}{Astrophys. Space Sci. Libr.}
  \textbf{\bibinfo{volume}{457}}, \bibinfo{pages}{455} (\bibinfo{year}{2018}),
  \eprint{1711.06520}.

\bibitem[{\citenamefont{Shternin et~al.}(2018)\citenamefont{Shternin, Baldo,
  and Haensel}}]{Shternin:2018dcn}
\bibinfo{author}{\bibfnamefont{P.~S.} \bibnamefont{Shternin}},
  \bibinfo{author}{\bibfnamefont{M.}~\bibnamefont{Baldo}}, \bibnamefont{and}
  \bibinfo{author}{\bibfnamefont{P.}~\bibnamefont{Haensel}},
  \bibinfo{journal}{Phys. Lett. B} \textbf{\bibinfo{volume}{786}},
  \bibinfo{pages}{28} (\bibinfo{year}{2018}), \eprint{1807.06569}.

\bibitem[{\citenamefont{Suleiman et~al.}(2023)\citenamefont{Suleiman, Oertel,
  and Mancini}}]{Suleiman:2023bdf}
\bibinfo{author}{\bibfnamefont{L.}~\bibnamefont{Suleiman}},
  \bibinfo{author}{\bibfnamefont{M.}~\bibnamefont{Oertel}}, \bibnamefont{and}
  \bibinfo{author}{\bibfnamefont{M.}~\bibnamefont{Mancini}},
  \bibinfo{journal}{Phys. Rev. C} \textbf{\bibinfo{volume}{108}},
  \bibinfo{pages}{035803} (\bibinfo{year}{2023}), \eprint{2308.09819}.

\bibitem[{\citenamefont{Roberts et~al.}(2012)\citenamefont{Roberts, Reddy, and
  Shen}}]{Roberts:2012um}
\bibinfo{author}{\bibfnamefont{L.~F.} \bibnamefont{Roberts}},
  \bibinfo{author}{\bibfnamefont{S.}~\bibnamefont{Reddy}}, \bibnamefont{and}
  \bibinfo{author}{\bibfnamefont{G.}~\bibnamefont{Shen}},
  \bibinfo{journal}{Phys. Rev. C} \textbf{\bibinfo{volume}{86}},
  \bibinfo{pages}{065803} (\bibinfo{year}{2012}), \eprint{1205.4066}.

\bibitem[{\citenamefont{Pascal et~al.}(2022)\citenamefont{Pascal, Novak, and
  Oertel}}]{Pascal:2022qeg}
\bibinfo{author}{\bibfnamefont{A.}~\bibnamefont{Pascal}},
  \bibinfo{author}{\bibfnamefont{J.}~\bibnamefont{Novak}}, \bibnamefont{and}
  \bibinfo{author}{\bibfnamefont{M.}~\bibnamefont{Oertel}},
  \bibinfo{journal}{Mon. Not. Roy. Astron. Soc.}
  \textbf{\bibinfo{volume}{511}}, \bibinfo{pages}{356} (\bibinfo{year}{2022}),
  \eprint{2201.01955}.

\bibitem[{\citenamefont{Burrows and Sawyer}(1999)}]{Burrows:1998ek}
\bibinfo{author}{\bibfnamefont{A.}~\bibnamefont{Burrows}} \bibnamefont{and}
  \bibinfo{author}{\bibfnamefont{R.~F.} \bibnamefont{Sawyer}},
  \bibinfo{journal}{Phys. Rev. C} \textbf{\bibinfo{volume}{59}},
  \bibinfo{pages}{510} (\bibinfo{year}{1999}), \eprint{astro-ph/9804264}.

\bibitem[{\citenamefont{Sedrakian and Dieperink}(2000)}]{Sedrakian:2000kc}
\bibinfo{author}{\bibfnamefont{A.}~\bibnamefont{Sedrakian}} \bibnamefont{and}
  \bibinfo{author}{\bibfnamefont{A.~E.~L.} \bibnamefont{Dieperink}},
  \bibinfo{journal}{Phys. Rev. D} \textbf{\bibinfo{volume}{62}},
  \bibinfo{pages}{083002} (\bibinfo{year}{2000}), \eprint{astro-ph/0002228}.

\bibitem[{\citenamefont{Lykasov et~al.}(2008)\citenamefont{Lykasov, Pethick,
  and Schwenk}}]{Lykasov:2008yz}
\bibinfo{author}{\bibfnamefont{G.~I.} \bibnamefont{Lykasov}},
  \bibinfo{author}{\bibfnamefont{C.~J.} \bibnamefont{Pethick}},
  \bibnamefont{and} \bibinfo{author}{\bibfnamefont{A.}~\bibnamefont{Schwenk}},
  \bibinfo{journal}{Phys. Rev. C} \textbf{\bibinfo{volume}{78}},
  \bibinfo{pages}{045803} (\bibinfo{year}{2008}), \eprint{0808.0330}.

\bibitem[{\citenamefont{Roberts and Reddy}(2017)}]{Roberts:2016mwj}
\bibinfo{author}{\bibfnamefont{L.~F.} \bibnamefont{Roberts}} \bibnamefont{and}
  \bibinfo{author}{\bibfnamefont{S.}~\bibnamefont{Reddy}},
  \bibinfo{journal}{Phys. Rev. C} \textbf{\bibinfo{volume}{95}},
  \bibinfo{pages}{045807} (\bibinfo{year}{2017}), \eprint{1612.02764}.

\bibitem[{\citenamefont{Vogel}(1984)}]{Vogel:1983hi}
\bibinfo{author}{\bibfnamefont{P.}~\bibnamefont{Vogel}},
  \bibinfo{journal}{Phys. Rev. D} \textbf{\bibinfo{volume}{29}},
  \bibinfo{pages}{1918} (\bibinfo{year}{1984}).

\bibitem[{\citenamefont{Horowitz and Perez-Garcia}(2003)}]{Horowitz:2003yx}
\bibinfo{author}{\bibfnamefont{C.~J.} \bibnamefont{Horowitz}} \bibnamefont{and}
  \bibinfo{author}{\bibfnamefont{M.~A.} \bibnamefont{Perez-Garcia}},
  \bibinfo{journal}{Phys. Rev. C} \textbf{\bibinfo{volume}{68}},
  \bibinfo{pages}{025803} (\bibinfo{year}{2003}), \eprint{astro-ph/0305138}.

\bibitem[{\citenamefont{Dieperink et~al.}(1990)\citenamefont{Dieperink,
  Piekarewicz, and Wehrberger}}]{Dieperink:1990kw}
\bibinfo{author}{\bibfnamefont{A.~E.~L.} \bibnamefont{Dieperink}},
  \bibinfo{author}{\bibfnamefont{J.}~\bibnamefont{Piekarewicz}},
  \bibnamefont{and}
  \bibinfo{author}{\bibfnamefont{K.}~\bibnamefont{Wehrberger}},
  \bibinfo{journal}{Phys. Rev. C} \textbf{\bibinfo{volume}{41}},
  \bibinfo{pages}{R2479} (\bibinfo{year}{1990}).

\bibitem[{\citenamefont{Peskin and Schroeder}(1995)}]{Peskin:1995ev}
\bibinfo{author}{\bibfnamefont{M.~E.} \bibnamefont{Peskin}} \bibnamefont{and}
  \bibinfo{author}{\bibfnamefont{D.~V.} \bibnamefont{Schroeder}},
  \emph{\bibinfo{title}{{An Introduction to quantum field theory}}}
  (\bibinfo{publisher}{Addison-Wesley}, \bibinfo{address}{Reading, USA},
  \bibinfo{year}{1995}), ISBN \bibinfo{isbn}{978-0-201-50397-5,
  978-0-429-50355-9, 978-0-429-49417-8}.

\bibitem[{\citenamefont{Leinson and Perez}(2001)}]{Leinson:2001ei}
\bibinfo{author}{\bibfnamefont{L.~B.} \bibnamefont{Leinson}} \bibnamefont{and}
  \bibinfo{author}{\bibfnamefont{A.}~\bibnamefont{Perez}},
  \bibinfo{journal}{Phys. Lett. B} \textbf{\bibinfo{volume}{518}},
  \bibinfo{pages}{15} (\bibinfo{year}{2001}), \bibinfo{note}{[Erratum:
  Phys.Lett.B 522, 358--358 (2001)]}, \eprint{hep-ph/0110207}.

\bibitem[{\citenamefont{Horowitz et~al.}(2012)\citenamefont{Horowitz, Shen,
  O'Connor, and Ott}}]{Horowitz:2012us}
\bibinfo{author}{\bibfnamefont{C.~J.} \bibnamefont{Horowitz}},
  \bibinfo{author}{\bibfnamefont{G.}~\bibnamefont{Shen}},
  \bibinfo{author}{\bibfnamefont{E.}~\bibnamefont{O'Connor}}, \bibnamefont{and}
  \bibinfo{author}{\bibfnamefont{C.~D.} \bibnamefont{Ott}},
  \bibinfo{journal}{Phys. Rev. C} \textbf{\bibinfo{volume}{86}},
  \bibinfo{pages}{065806} (\bibinfo{year}{2012}), \eprint{1209.3173}.

\bibitem[{\citenamefont{Kuksa}(2015)}]{Kuksa:2015}
\bibinfo{author}{\bibfnamefont{V.}~\bibnamefont{Kuksa}},
  \bibinfo{journal}{Advances in High Energy Physics}
  \textbf{\bibinfo{volume}{2015}}, \bibinfo{pages}{490238}
  (\bibinfo{year}{2015}), \eprint{1408.6994}.

\bibitem[{\citenamefont{Fattoyev et~al.}(2010)\citenamefont{Fattoyev, Horowitz,
  Piekarewicz, and Shen}}]{Fattoyev:2010mx}
\bibinfo{author}{\bibfnamefont{F.~J.} \bibnamefont{Fattoyev}},
  \bibinfo{author}{\bibfnamefont{C.~J.} \bibnamefont{Horowitz}},
  \bibinfo{author}{\bibfnamefont{J.}~\bibnamefont{Piekarewicz}},
  \bibnamefont{and} \bibinfo{author}{\bibfnamefont{G.}~\bibnamefont{Shen}},
  \bibinfo{journal}{Phys. Rev. C} \textbf{\bibinfo{volume}{82}},
  \bibinfo{pages}{055803} (\bibinfo{year}{2010}), \eprint{1008.3030}.

\bibitem[{\citenamefont{Alford and Harris}(2018)}]{Alford:2018lhf}
\bibinfo{author}{\bibfnamefont{M.~G.} \bibnamefont{Alford}} \bibnamefont{and}
  \bibinfo{author}{\bibfnamefont{S.~P.} \bibnamefont{Harris}},
  \bibinfo{journal}{Phys. Rev. C} \textbf{\bibinfo{volume}{98}},
  \bibinfo{pages}{065806} (\bibinfo{year}{2018}), \eprint{1803.00662}.

\bibitem[{\citenamefont{Alford et~al.}(2021{\natexlab{b}})\citenamefont{Alford,
  Haber, Harris, and Zhang}}]{Alford:2021ogv}
\bibinfo{author}{\bibfnamefont{M.~G.} \bibnamefont{Alford}},
  \bibinfo{author}{\bibfnamefont{A.}~\bibnamefont{Haber}},
  \bibinfo{author}{\bibfnamefont{S.~P.} \bibnamefont{Harris}},
  \bibnamefont{and} \bibinfo{author}{\bibfnamefont{Z.}~\bibnamefont{Zhang}},
  \bibinfo{journal}{Universe} \textbf{\bibinfo{volume}{7}},
  \bibinfo{pages}{399} (\bibinfo{year}{2021}{\natexlab{b}}),
  \eprint{2108.03324}.

\bibitem[{\citenamefont{Gavassino et~al.}(2021)\citenamefont{Gavassino,
  Antonelli, and Haskell}}]{Gavassino:2020kwo}
\bibinfo{author}{\bibfnamefont{L.}~\bibnamefont{Gavassino}},
  \bibinfo{author}{\bibfnamefont{M.}~\bibnamefont{Antonelli}},
  \bibnamefont{and} \bibinfo{author}{\bibfnamefont{B.}~\bibnamefont{Haskell}},
  \bibinfo{journal}{Class. Quant. Grav.} \textbf{\bibinfo{volume}{38}},
  \bibinfo{pages}{075001} (\bibinfo{year}{2021}), \eprint{2003.04609}.

\bibitem[{\citenamefont{Radice et~al.}(2022)\citenamefont{Radice, Bernuzzi,
  Perego, and Haas}}]{Radice:2021jtw}
\bibinfo{author}{\bibfnamefont{D.}~\bibnamefont{Radice}},
  \bibinfo{author}{\bibfnamefont{S.}~\bibnamefont{Bernuzzi}},
  \bibinfo{author}{\bibfnamefont{A.}~\bibnamefont{Perego}}, \bibnamefont{and}
  \bibinfo{author}{\bibfnamefont{R.}~\bibnamefont{Haas}},
  \bibinfo{journal}{Mon. Not. Roy. Astron. Soc.}
  \textbf{\bibinfo{volume}{512}}, \bibinfo{pages}{1499} (\bibinfo{year}{2022}),
  \eprint{2111.14858}.

\bibitem[{\citenamefont{Gavassino and Noronha}(2024)}]{Gavassino:2023xkt}
\bibinfo{author}{\bibfnamefont{L.}~\bibnamefont{Gavassino}} \bibnamefont{and}
  \bibinfo{author}{\bibfnamefont{J.}~\bibnamefont{Noronha}},
  \bibinfo{journal}{Phys. Rev. D} \textbf{\bibinfo{volume}{109}},
  \bibinfo{pages}{096040} (\bibinfo{year}{2024}), \eprint{2305.04119}.

\bibitem[{\citenamefont{Fortin et~al.}(2018)\citenamefont{Fortin, Taranto,
  Burgio, Haensel, Schulze, and Zdunik}}]{Fortin:2017rxq}
\bibinfo{author}{\bibfnamefont{M.}~\bibnamefont{Fortin}},
  \bibinfo{author}{\bibfnamefont{G.}~\bibnamefont{Taranto}},
  \bibinfo{author}{\bibfnamefont{G.~F.} \bibnamefont{Burgio}},
  \bibinfo{author}{\bibfnamefont{P.}~\bibnamefont{Haensel}},
  \bibinfo{author}{\bibfnamefont{H.~J.} \bibnamefont{Schulze}},
  \bibnamefont{and} \bibinfo{author}{\bibfnamefont{J.~L.}
  \bibnamefont{Zdunik}}, \bibinfo{journal}{Mon. Not. Roy. Astron. Soc.}
  \textbf{\bibinfo{volume}{475}}, \bibinfo{pages}{5010} (\bibinfo{year}{2018}),
  \eprint{1709.04855}.

\bibitem[{\citenamefont{Fornal and Grinstein}(2018)}]{Fornal:2018eol}
\bibinfo{author}{\bibfnamefont{B.}~\bibnamefont{Fornal}} \bibnamefont{and}
  \bibinfo{author}{\bibfnamefont{B.}~\bibnamefont{Grinstein}},
  \bibinfo{journal}{Phys. Rev. Lett.} \textbf{\bibinfo{volume}{120}},
  \bibinfo{pages}{191801} (\bibinfo{year}{2018}), \bibinfo{note}{[Erratum:
  Phys.Rev.Lett. 124, 219901 (2020)]}, \eprint{1801.01124}.

\bibitem[{\citenamefont{Husain et~al.}(2022)\citenamefont{Husain, Motta, and
  Thomas}}]{Husain:2022bxl}
\bibinfo{author}{\bibfnamefont{W.}~\bibnamefont{Husain}},
  \bibinfo{author}{\bibfnamefont{T.~F.} \bibnamefont{Motta}}, \bibnamefont{and}
  \bibinfo{author}{\bibfnamefont{A.~W.} \bibnamefont{Thomas}},
  \bibinfo{journal}{JCAP} \textbf{\bibinfo{volume}{10}}, \bibinfo{pages}{028}
  (\bibinfo{year}{2022}), \eprint{2203.02758}.

\bibitem[{\citenamefont{Shirke et~al.}(2023)\citenamefont{Shirke, Ghosh,
  Chatterjee, Sagunski, and Schaffner-Bielich}}]{Shirke:2023ktu}
\bibinfo{author}{\bibfnamefont{S.}~\bibnamefont{Shirke}},
  \bibinfo{author}{\bibfnamefont{S.}~\bibnamefont{Ghosh}},
  \bibinfo{author}{\bibfnamefont{D.}~\bibnamefont{Chatterjee}},
  \bibinfo{author}{\bibfnamefont{L.}~\bibnamefont{Sagunski}}, \bibnamefont{and}
  \bibinfo{author}{\bibfnamefont{J.}~\bibnamefont{Schaffner-Bielich}},
  \bibinfo{journal}{JCAP} \textbf{\bibinfo{volume}{12}}, \bibinfo{pages}{008}
  (\bibinfo{year}{2023}), \eprint{2305.05664}.

\bibitem[{\citenamefont{Shirke et~al.}(2024)\citenamefont{Shirke, Pradhan,
  Chatterjee, Sagunski, and Schaffner-Bielich}}]{Shirke:2024ymc}
\bibinfo{author}{\bibfnamefont{S.}~\bibnamefont{Shirke}},
  \bibinfo{author}{\bibfnamefont{B.~K.} \bibnamefont{Pradhan}},
  \bibinfo{author}{\bibfnamefont{D.}~\bibnamefont{Chatterjee}},
  \bibinfo{author}{\bibfnamefont{L.}~\bibnamefont{Sagunski}}, \bibnamefont{and}
  \bibinfo{author}{\bibfnamefont{J.}~\bibnamefont{Schaffner-Bielich}}
  (\bibinfo{year}{2024}), \eprint{2403.18740}.

\bibitem[{\citenamefont{Haensel et~al.}(2002)\citenamefont{Haensel, Levenfish,
  and Yakovlev}}]{Haensel:2001em}
\bibinfo{author}{\bibfnamefont{P.}~\bibnamefont{Haensel}},
  \bibinfo{author}{\bibfnamefont{K.~P.} \bibnamefont{Levenfish}},
  \bibnamefont{and} \bibinfo{author}{\bibfnamefont{D.~G.}
  \bibnamefont{Yakovlev}}, \bibinfo{journal}{Astron. Astrophys.}
  \textbf{\bibinfo{volume}{381}}, \bibinfo{pages}{1080} (\bibinfo{year}{2002}),
  \eprint{astro-ph/0110575}.

\bibitem[{\citenamefont{van Dalen and Dieperink}(2004)}]{vanDalen:2003uy}
\bibinfo{author}{\bibfnamefont{E.~N.~E.} \bibnamefont{van Dalen}}
  \bibnamefont{and} \bibinfo{author}{\bibfnamefont{A.~E.~L.}
  \bibnamefont{Dieperink}}, \bibinfo{journal}{Phys. Rev. C}
  \textbf{\bibinfo{volume}{69}}, \bibinfo{pages}{025802}
  (\bibinfo{year}{2004}), \eprint{nucl-th/0311103}.

\bibitem[{\citenamefont{Ofengeim et~al.}(2019)\citenamefont{Ofengeim, Gusakov,
  Haensel, and Fortin}}]{Ofengeim:2019fjy}
\bibinfo{author}{\bibfnamefont{D.~D.} \bibnamefont{Ofengeim}},
  \bibinfo{author}{\bibfnamefont{M.~E.} \bibnamefont{Gusakov}},
  \bibinfo{author}{\bibfnamefont{P.}~\bibnamefont{Haensel}}, \bibnamefont{and}
  \bibinfo{author}{\bibfnamefont{M.}~\bibnamefont{Fortin}},
  \bibinfo{journal}{Phys. Rev. D} \textbf{\bibinfo{volume}{100}},
  \bibinfo{pages}{103017} (\bibinfo{year}{2019}), \eprint{1911.08407}.

\bibitem[{\citenamefont{Alford and Haber}(2021)}]{Alford:2020pld}
\bibinfo{author}{\bibfnamefont{M.~G.} \bibnamefont{Alford}} \bibnamefont{and}
  \bibinfo{author}{\bibfnamefont{A.}~\bibnamefont{Haber}},
  \bibinfo{journal}{Phys. Rev. C} \textbf{\bibinfo{volume}{103}},
  \bibinfo{pages}{045810} (\bibinfo{year}{2021}), \eprint{2009.05181}.

\bibitem[{\citenamefont{Xu et~al.}(2021)\citenamefont{Xu, Huang, Wang, Liu, and
  Han}}]{Xu:2021jns}
\bibinfo{author}{\bibfnamefont{Y.}~\bibnamefont{Xu}},
  \bibinfo{author}{\bibfnamefont{X.~L.} \bibnamefont{Huang}},
  \bibinfo{author}{\bibfnamefont{Y.~B.} \bibnamefont{Wang}},
  \bibinfo{author}{\bibfnamefont{C.~Z.} \bibnamefont{Liu}}, \bibnamefont{and}
  \bibinfo{author}{\bibfnamefont{J.~L.} \bibnamefont{Han}},
  \bibinfo{journal}{Int. J. Mod. Phys. D} \textbf{\bibinfo{volume}{30}},
  \bibinfo{pages}{2150054} (\bibinfo{year}{2021}).

\bibitem[{\citenamefont{Jaikumar et~al.}(2006)\citenamefont{Jaikumar, Roberts,
  and Sedrakian}}]{Jaikumar:2005hy}
\bibinfo{author}{\bibfnamefont{P.}~\bibnamefont{Jaikumar}},
  \bibinfo{author}{\bibfnamefont{C.~D.} \bibnamefont{Roberts}},
  \bibnamefont{and}
  \bibinfo{author}{\bibfnamefont{A.}~\bibnamefont{Sedrakian}},
  \bibinfo{journal}{Phys. Rev. C} \textbf{\bibinfo{volume}{73}},
  \bibinfo{pages}{042801} (\bibinfo{year}{2006}), \eprint{nucl-th/0509093}.

\bibitem[{\citenamefont{Berdermann et~al.}(2016)\citenamefont{Berdermann,
  Blaschke, Fischer, and Kachanovich}}]{Berdermann:2016mwt}
\bibinfo{author}{\bibfnamefont{J.}~\bibnamefont{Berdermann}},
  \bibinfo{author}{\bibfnamefont{D.}~\bibnamefont{Blaschke}},
  \bibinfo{author}{\bibfnamefont{T.}~\bibnamefont{Fischer}}, \bibnamefont{and}
  \bibinfo{author}{\bibfnamefont{A.}~\bibnamefont{Kachanovich}},
  \bibinfo{journal}{Phys. Rev. D} \textbf{\bibinfo{volume}{94}},
  \bibinfo{pages}{123010} (\bibinfo{year}{2016}), \eprint{1609.05201}.

\bibitem[{\citenamefont{Hernandez et~al.}(2024)\citenamefont{Hernandez, Manuel,
  and Tolos}}]{Hernandez:2024rxi}
\bibinfo{author}{\bibfnamefont{J.~L.} \bibnamefont{Hernandez}},
  \bibinfo{author}{\bibfnamefont{C.}~\bibnamefont{Manuel}}, \bibnamefont{and}
  \bibinfo{author}{\bibfnamefont{L.}~\bibnamefont{Tolos}},
  \bibinfo{journal}{Phys. Rev. D} \textbf{\bibinfo{volume}{109}},
  \bibinfo{pages}{123022} (\bibinfo{year}{2024}), \eprint{2402.06595}.

\bibitem[{\citenamefont{Vidana et~al.}(2022)\citenamefont{Vidana, Logoteta, and
  Bombaci}}]{Vidana:2022ket}
\bibinfo{author}{\bibfnamefont{I.}~\bibnamefont{Vidana}},
  \bibinfo{author}{\bibfnamefont{D.}~\bibnamefont{Logoteta}}, \bibnamefont{and}
  \bibinfo{author}{\bibfnamefont{I.}~\bibnamefont{Bombaci}},
  \bibinfo{journal}{Phys. Rev. C} \textbf{\bibinfo{volume}{106}},
  \bibinfo{pages}{035804} (\bibinfo{year}{2022}), \eprint{2206.10190}.

\bibitem[{\citenamefont{Shin et~al.}(2024)\citenamefont{Shin, Rrapaj, Holt, and
  Reddy}}]{Shin:2023sei}
\bibinfo{author}{\bibfnamefont{E.}~\bibnamefont{Shin}},
  \bibinfo{author}{\bibfnamefont{E.}~\bibnamefont{Rrapaj}},
  \bibinfo{author}{\bibfnamefont{J.~W.} \bibnamefont{Holt}}, \bibnamefont{and}
  \bibinfo{author}{\bibfnamefont{S.~K.} \bibnamefont{Reddy}},
  \bibinfo{journal}{Phys. Rev. C} \textbf{\bibinfo{volume}{109}},
  \bibinfo{pages}{015804} (\bibinfo{year}{2024}), \eprint{2306.05280}.

\end{thebibliography}

\end{document}